\documentclass[aps,pra,twocolumn,showpacs,amsmath,amssymb,superscriptaddress,longbibliography]{revtex4-1}

\usepackage{bm}
\usepackage{amsthm}
\usepackage{amsfonts}
\usepackage{graphicx}
\usepackage{epsfig}
\usepackage{epstopdf}
\usepackage{subfigure}
\usepackage[colorlinks=true,citecolor=blue,urlcolor=blue]{hyperref}
\usepackage{color}
\usepackage{ulem}

\begin{document}

\title{Topological transition from superfluid vortex rings to isolated knots and links}

\author{Wen-Kai Bai}
\affiliation{Shaanxi Key Laboratory for Theoretical Physics Frontiers,  Institute of Modern Physics, Northwest University, Xi'an 710127, China}
\author{Tao Yang}
\email{yangt@nwu.edu.cn}
\affiliation{Shaanxi Key Laboratory for Theoretical Physics Frontiers,  Institute of Modern Physics, Northwest University, Xi'an 710127, China}
\affiliation{School of Physics, Northwest University, Xi'an 710127, China}
\affiliation{NSFC-SPTP Peng Huanwu Center for Fundamental Theory, Xian 710127, China}

\author{Wu-Ming Liu}
\email{wmliu@iphy.ac.cn}
\affiliation{Beijing National Laboratory for Condensed Matter Physics, Institute of Physics, Chinese Academy of Sciences, Beijing, China}
\affiliation{School of Physical Sciences, University of Chinese Academy of Sciences, Beijing 100190, China}
\affiliation{Songshan Lake Materials Laboratory, Dongguan, Guangdong 523808, China}

\date{\today}

\begin{abstract}
Knots and links are fundamental topological objects play a key role in both classical and quantum fluids. In this research, we propose a novel scheme to generate torus vortex knots and links through the reconnections of vortex rings perturbed by Kelvin waves in trapped Bose-Einstein condensates. We observe a new phenomenon in a confined superfluid system in which the transfer of helicity between knots/links and coils can occur in both directions with different pathways. The pathways of topology transition can be controlled through designing specific initial states. The generation of a knot or link can be achieved by setting the parity of the Kelvin wave number. The stability of knots/links can be improved greatly with tunable parameters, including the ideal relative angle and the minimal distance between the initial vortex rings.
\end{abstract}

\maketitle

\section{Introduction}

Knots and links are of great interest in many areas of science including physics, chemistry and biology. Although they commonly form in nature, everyday occurrences we have all experienced, it is difficult to achieve the controllable formation of these structures with complex topologies. Experimental and theoretical observations of knots and links exhibit highly divergent and nonlinear dynamics, which is vital for understanding various persistent phenomena and turbulent behaviors, ranging from water\,\cite{nphys.9.253,PNAS.111.15350}, superfluid systems\,\cite{nphys.5.193,nphys.12.650,PNAS.111.15350,nphys.12.478,NJP.18.063016,PRE.85.036306}, plasma\,\cite{PNAS.111.3663,phys.tod.49.28}, agitated strings\,\cite{PNAS.104.16432}, to liquid crystals\,\cite{Science.333.62,nmat.13.258}, and with increasing importance in a variety of scenarios, such as synthesising DNA/RNA in biological systems\,\cite{Science.232.951,science.257.1110,nnano.5.712,Science.304.1308,NAMS.42.528,PNAS.110.20906,ncomm.8.14936}, and molecular designing in chemistry\,\cite{Chem.Rev.111.5434,science.352.1555,science.355.159,Science.365.272}. In quantum fluids, knots and links are the tangled filaments of vortices. Ultra-cold atomic Bose-Einstein condensates (BECs) provide a controllable platform for both comprehensive theoretical studies of these topological excitations and direct observations of their dynamics using tunable parameters. The discrete filamentary nature of vortices is an advantage of quantum fluids in studies of vortex interactions and reconnections over ordinary fluids. Additionally, this characteristic is helpful for understanding behaviours, such as various persistent phenomena and turbulence, in viscous classical fluid and biological systems due to similarities discovered in previous research\,\cite{nphys.9.253,Science.357.487,PNAS.111.15350,Science.232.951,nphys.12.650}.

Great efforts have been made to create knots and links in different classical contexts. For example, isolated trefoil vortex knots and pairs of linked vortex rings were created in water using a method of accelerating specially shaped hydrofoils produced by 3D printing technology\,\cite{nphys.9.253}, isolated optical vortex loops in the forms of knots and links were realized by optical beams with using algebraic topology\,\cite{nphys.6.118}, and many others including those in biology and chemistry. However, the generation process of knots and links remains challenging in quantum fluids. Further study has been hindered by the lack of techniques to generate knots/links. Most of studies of fluid systems and the DNA replication process have been mainly performed in homogeneous systems, away from boundaries, to study the dissolving of knots/links\,\cite{nphys.12.650,PNAS.110.20906,PNAS.111.15350,srep.6.24118,srep.7.12420,PRE.95.053109, PRE.85.036306} or head-on collisions\,\cite{nature.357.225,PRL.76.4745}, where only standard reconnections occurs. In this process, the reduced complexity of the topological structures is identified through stepwise reduction, which show the transfer of helicity of the system evolves only in one direction from knotted/linked structures to helical coils. A recent study\,\cite{PRX.7.021031} observed the boundary effects on vortex dynamics that allow double reconnections, rebounds, and ejections of vortex lines in a cigar-shaped atomic BEC in addition to standard reconnections in infinite uniform fluids. Changes may also occur in the dynamics of vortex structures with complex topologies in confined systems. Experimentally, condensates are confined in optomagnetic traps, which are suitable for exploring confined systems with different geometries.

During the reconnection process, cusps are generated, and the vortex lines are excited by Kelvin waves, helical perturbations that travel along the vortices, as they relax\,\cite{PRL.86.3080,PRX.7.021031}. In turn, the Kelvin waves provide a source of perturbations for the motion of vortices. The Kelvin wave cascade generated in superfluid turbulence reflects the importance of Kelvin waves and reconnections in the transfer of energy\,\cite{PRL.86.3080}, which may be exploited to stabilise vortex structures. The knots/links dynamics with Kelvin perturbations in confined systems remains a topic of research. The objective of our study is to make the reconnections of vortices occur controllably and then form the complex structures wanted in confined BECs.

\section{Model for Numerical simulations}

To achieve the goal of generating knots/links, 
we use the topologically trivial objects, that is, vortex rings in a spherical BEC, as building blocks. A vortex ring is a stable nonlinear excitation mode that can be simply described as a vortex line that has been bent into a closed loop. In a trapped BEC, a vortex ring moves in response to the effect of the nonuniform trap potential and the external rotation, in addition to self-induced effects caused by the local curvature of the ring\,\cite{JPCM.13.R135}. By introducing Kelvin waves into the system, the vortex ring is modified by periodic distortions, which reduces the translational self-induced velocity of the vortex ring\,\cite{PRE.74.046303,PRA.83.045601}, and adds helicity to the vortex ring. In ideal fluids, which lack viscosity, helicity\,\cite{JFM.35.117} is a conserved quantity that measures the degree of knottedness and entanglement of a fluid flow. Helicity can be stored by twisting, writhe (coiling and knotting), and linking\,\cite{phys.tod.49.28}. In quantum fluids, helicity varies with topology-changing reconnections and changes of geometry in helical vortices. We discuss the evolution of helicity in detail in Supplementary Information.

The interaction between vortices is another factor that plays an important role in vortex dynamics and leads to interesting phenomena. In homogeneous systems, one of the most impressive sights occurs when two same-sized vortex rings placed front to back moves together in the same direction by leapfrogging through each other\,\cite{PRL.94.124502}. When one of the rings is helically wound initially, the centreline helicity varies considerably, and the leapfrogging motion occurs due to the stretching and compressing of the helical ring\,\cite{Science.357.487,PNAS.111.15350}. However, the reconnections do not occur between these independent rings.

At sufficiently low temperatures, the macroscopic behavior of a trapped BEC with $N$ atoms is well characterized by the Gross-Pitaevskii (GP) equation, which is useful for studying topological vortex excitations:
\begin{align}
i\hbar\frac{\partial \psi}{\partial t}=\left(-\frac{\hbar^2}{2m}\nabla^2+V_{tr}\left(\mathbf{r}\right)+g\left|\psi\right|^2\right)\psi\text{,}
\label{GPE}
\end{align}
where $\psi$ is the wave function, the coupling constant $g=4\pi\hbar^2a_s/m$ is related to the $s$-wave scattering length $a_s$ of the atoms, and $m$ is the mass of the atoms. In this study, we use an isotropic harmonic trap $V_{tr}\left(\mathbf{r}\right)=m\omega^2\mathbf{r}^2/2$, where $\omega$ is the trap frequency. The oscillation length of the trap and unit of time are $a_0=\sqrt{\hbar/m\omega}$ and $t_0=1/\omega$, respectively.

In our simulations, the isotropic trap frequency is $\omega=2\pi\times 75 ~\text{Hz}$. The bulk s-wave scattering length $a_s = 5.4 \,\text{nm}$ and the mass $m=1.443\times 10^{-25}\,\text{kg}$ for the {$^{87}$\it{$Rb$}} BEC. We use $a_0=\sqrt{\hbar/m\omega}$ and $t_0=1/\omega$ as the units of length and time, respectively. In the calculations, $151\times 151\times 151$ grids with steps $\Delta_x=\Delta_y=\Delta_z=\sqrt{2}a_0/10$ are used in a uniformly discretised physical space. A small time step, $\Delta_ t=0.001t_0$, is chosen to ensure the accuracy of the results. The temporal evolution of the superfluid order parameter was computed by numerically integrating the Gross-Pitaevskii equation using the Crank-Nicolson method and fourth-order Ronger-Kutta method, which give quantitative agreement.

In a spherically trapped condensate, two unperturbed vortex rings of similar size can undergo leapfrogging motion back and forth without reconnections. This process occurs when only one vortex ring is perturbed by Kelvin waves. In a trapped condensate, when the minimum distance between the vortices reaches the order of the Thomas-Fermi radius $R_{TF}$ for the shortest direction in the condensate, the vortices start to rotate to arrange themselves in an antiparallel orientation, followed by reconnection in a direction orthogonal to their mutual alignment before final separation\,\cite{srep.5.9224,PRX.7.021031}. This process is a general characteristic of reconnection events, even in DNA biology\,\cite{NAMS.42.528, Maths.Proc.Camb.Phil.Soc.136.565}. The initial relative orientation and velocity of the vortex lines are crucial factors that influence the evolution of the vortices in a trapped condensate\,\cite{PRX.7.021031, PNAS.116.12204}. The objective of the study is to create knots or links of different topological types in a trapped condensate. Thus, we begin with two separated vortex rings perturbed by helical Kelvin waves of a given amplitude and azimuthal wave number to ensure that reconnections occur.

We label links and knots using the generalized method proposed by Scott and Dror called `Knot Atlas' (http://katlas.org). The first topologically non-trivial knot is the trefoil knot K3-1 with a crossing number and topological writhe of $n_c=3$ and $n_w=3$, respectively. For links, the second non-trivial link is Solomon link L4a1 with $n_c=n_w=4$. In the following paragraph, we use these two topological structures as examples to describe our scheme for generating knots and links with different topologies.

\begin{figure*}[tbhp]
\centering
\includegraphics[width=1.0\linewidth]{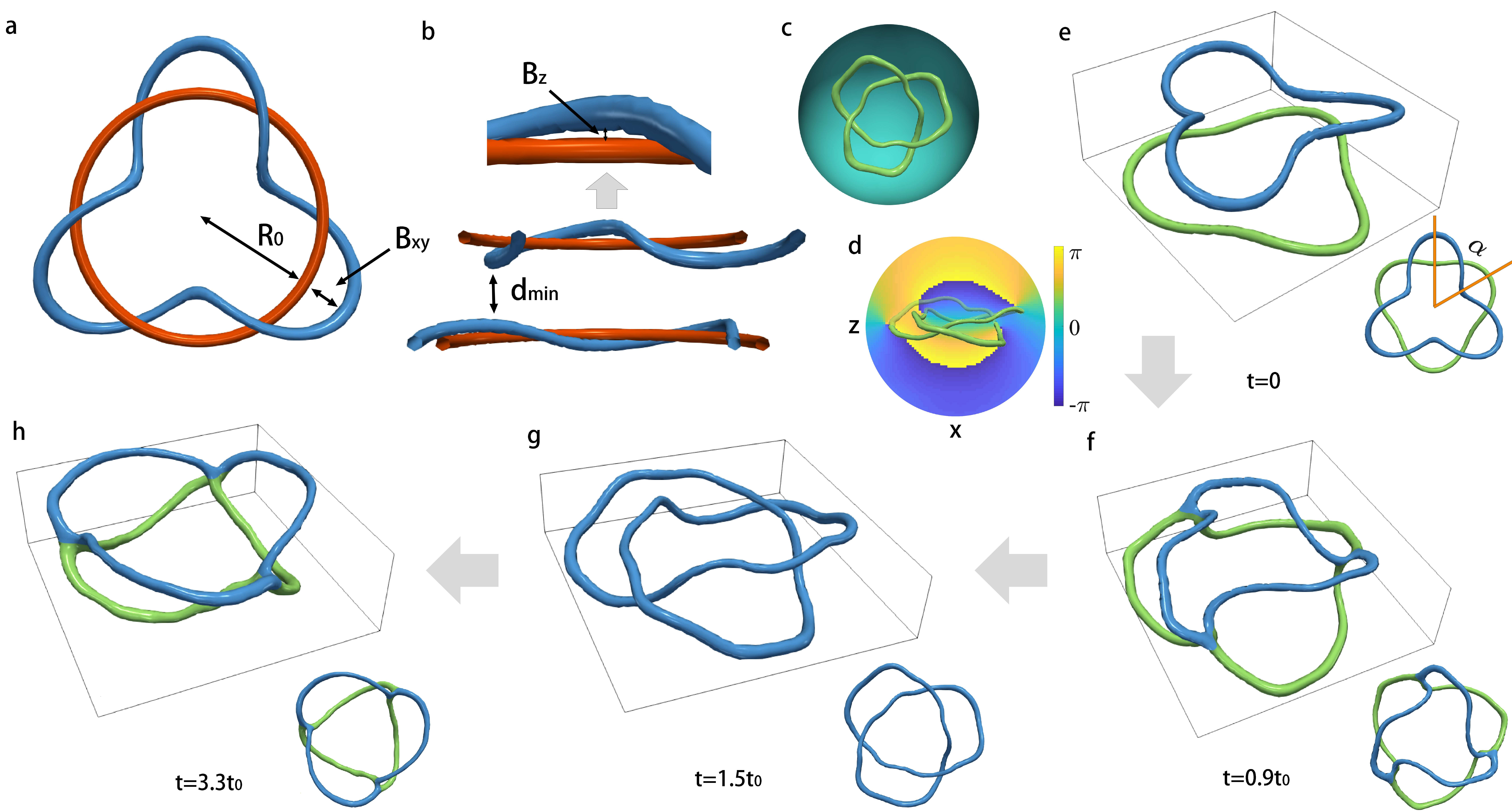}
\caption{{Initial state and geometric evolution from vortex rings to a trefoil vortex knot.} \textbf{a}, \textbf{b} Schematic diagram of a vortex ring perturbed by Kelvin waves with wave number $n=3$. The unperturbed red ring is used as a reference. \textbf{c}, \textbf{d}, Example of the density isosurface of a trefoil vortex knot in a spherical condensate and a cross section of the corresponding phase distribution. \textbf{e-h}, Four successive typical snapshots that show how two distorted vortex rings tie into a K3-1 trefoil vortex knot and then break up again. The trefoil vortex knot emerges at $1.0t_0$ and decays at $3.2t_0$. The initial radii and the minimal distance of the upper and lower vortex rings are $R_1=2.0a_0$, $R_2=2.3a_0$, and $d_{min}=1.0a_0$, respectively. The amplitudes of the Kelvin waves for the upper ring are $B_{xy1}=0.8a_0$ and $B_{z1}=0.3a_0$, and for the lower ring, $B_{xy2}=0.4a_0$ and $B_{z2}=0.2a_0$. The initial relative angle between the rings is $\alpha = \pi/3$, as shown in (e).}
\label{fig:k31}
\end{figure*}

\begin{figure*}[tbhp]
	\centering
	\includegraphics[width=1.0\textwidth]{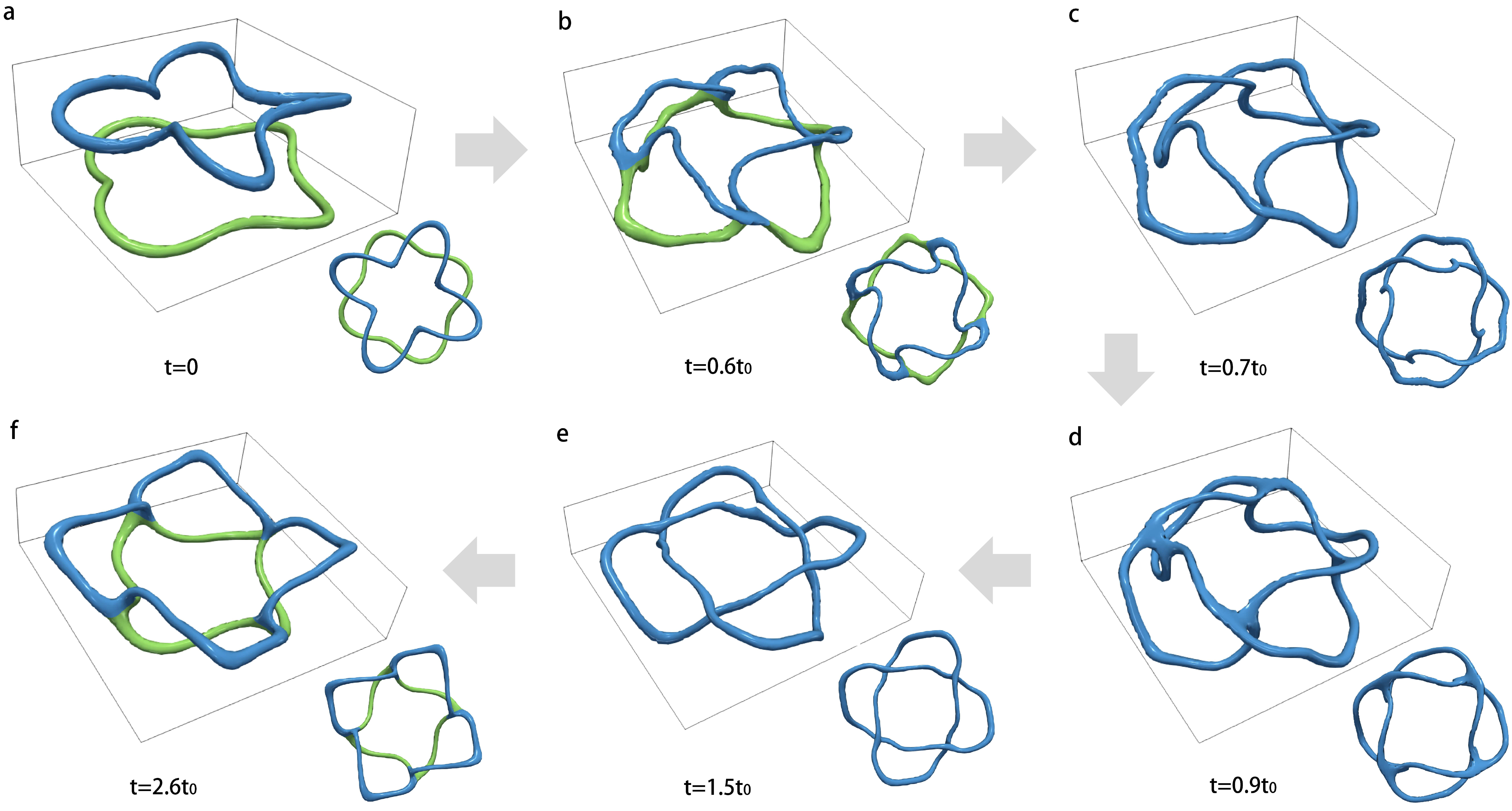}
	\caption{{Generation of a L4a1 Solomon vortex link.} \textbf{a-f}, Six successive snapshots that show how two perturbed vortex rings with Kelvin wave number $n=4$ evolve into the L4a1 vortex link and then break up again. The link appears twice between $0.7t_0$ and $0.8t_0$ (\textbf{c}) and between $1.1t_0$ and $2.5t_0$ (\textbf{e}). All other parameters are the same as those in Fig.\,\ref{fig:k31}.
}
\label{fig:l4a1}
\end{figure*}


\begin{figure*}[t]
	\centering
	\includegraphics[width=0.95\textwidth]{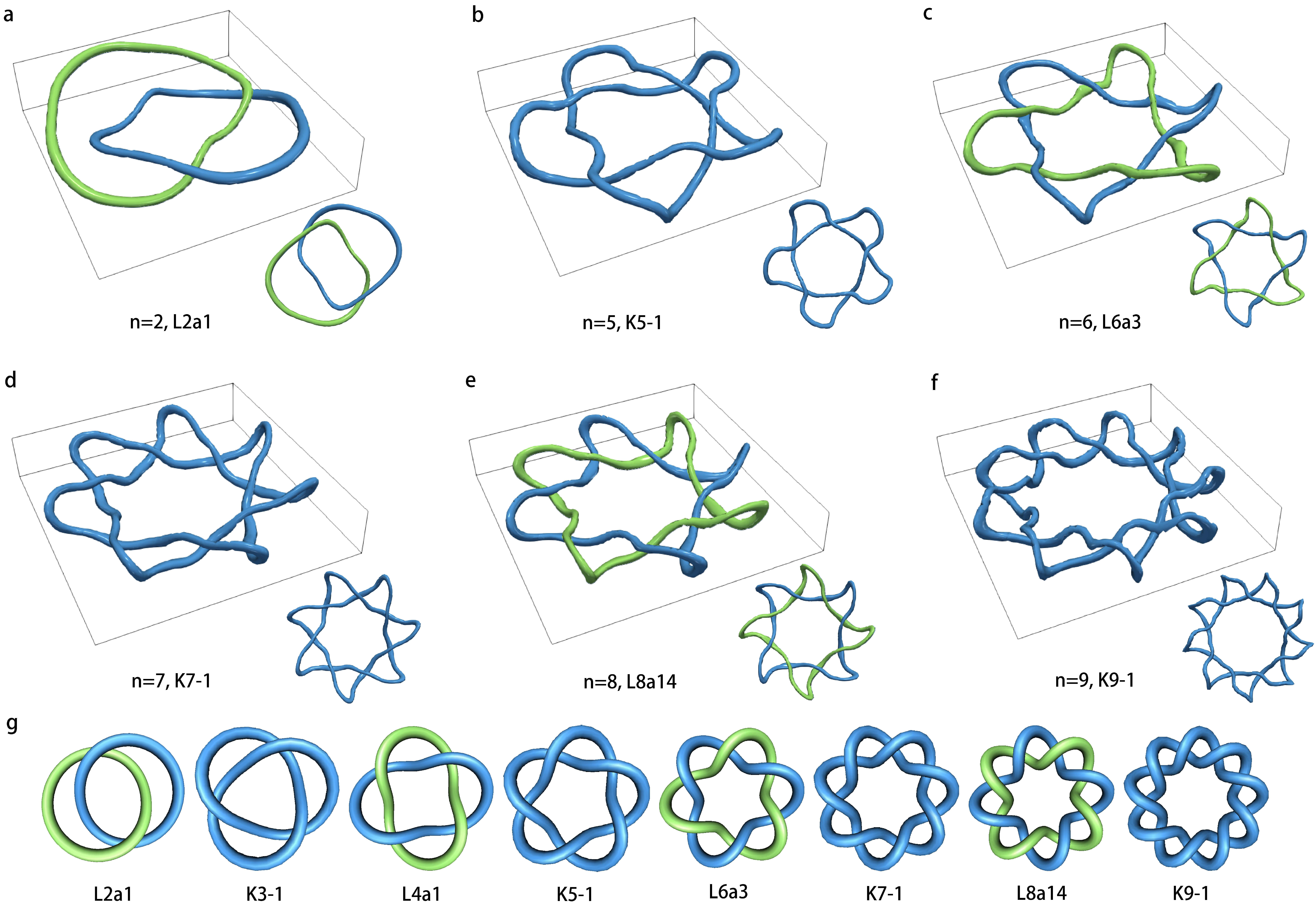}
	\caption{{Torus vortex links/knots generated with crossing number up to $n=9$.} \textbf{a} L2a1 link at $2.0t_0$. \textbf{b} K5-1 knot at $1.5t_0$. \textbf{c} L6a3 link at $t=1.5t_0$. \textbf{d} K7-1 knot at $t=1.5t_0$. \textbf{e} L8a14 link at $1.2t_0$. \textbf{f} K9-1 knot at $1.0t_0$. \textbf{g} Standard torus links and knots with crossing number up to 9 for KnotPlot Software configurations. All parameters are the same as those in Fig.\,\ref{fig:k31}, except for the Kelvin wave number, which equals the crossing number of the knots or links.}
	\label{fig:generalizeresult}
\end{figure*}

\begin{figure*}[t]
	\centering
	\includegraphics[width=\textwidth]{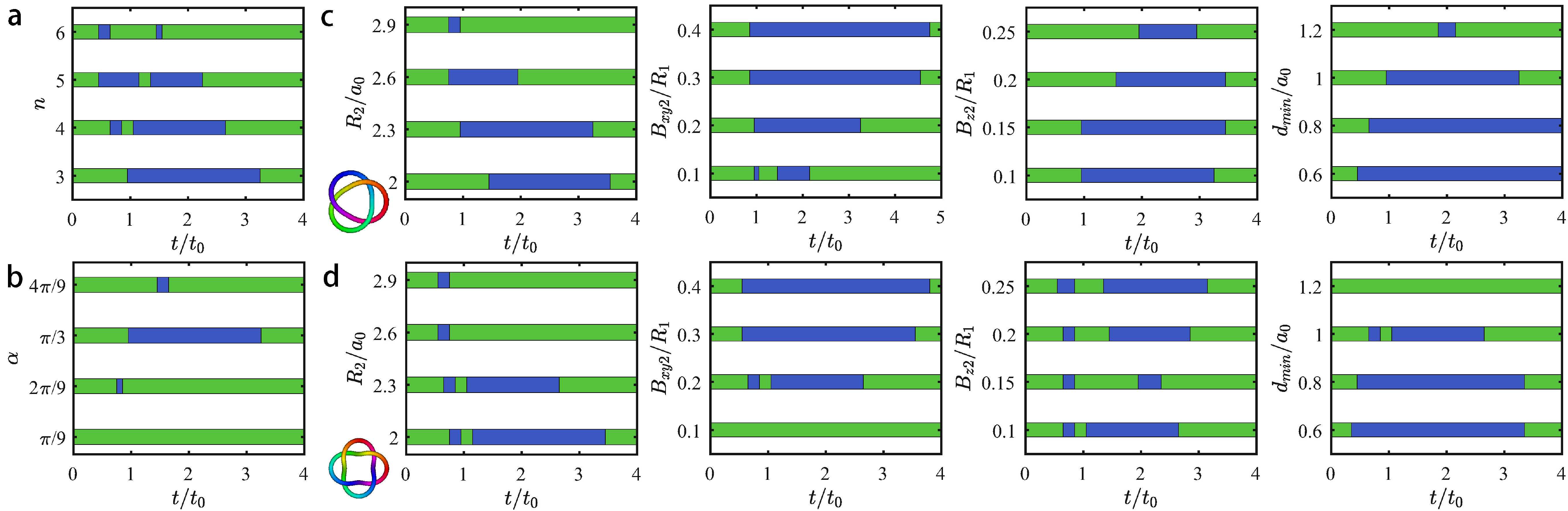}
	\caption{{Stability analysis of knots/links.} \textbf{a-b} Lifetime of a knot or link as a function of the Kelvin wave number $n$ and the relative angle $\alpha$ between the vortex rings.  \textbf{c} From left to right are the lifetime of trefoil knot K3-1 as a function of the radius of the lower ring $R_2$, the radial amplitude of the Kelvin waves of the lower ring $B_{xy2}$, the axial amplitude of the Kelvin waves of the lower ring $B_{z2}$, and the initial offset between the two rings $d_{min}$. \textbf{d} The same dependent parameters as in (b) for the lifetime of the Solomon link L4a1. In all plots, the initial parameters for the upper ring are the same as those in Fig.\,\ref{fig:k31}.}
	\label{fig:stability}
\end{figure*}

A vortex ring is formed by a loop in a vortex line. Suppose that a vortex ring is initially placed on the $xOy$ plane and symmetrically rotated about the $z$-axis; in this case, it can be considered as an assembly of 2D vortex dipoles in the $rOz$ plane. A sufficiently accurate description of a two-dimensional vortex centred at the origin of the $rOz$ plane is given by the wave function $\psi_{2D}\left(r,z\right)=\sqrt{\rho\left(l\right)}\exp{[i\theta\left(r,z\right)]}$, with $\rho\left(l\right)=l^2\left(a_1+a_2l^2\right)/\left(1+b_1l^2+b_2l^4\right)$, $\theta\left(r,z\right)={\rm {atan2}}\left(z,r\right)$, $r=\sqrt{x^2+y^2}$, and $l=\sqrt{r^2+z^2}$, where $a_j$ and $b_j$ are constants, and ${\rm {atan2}}(...)$ is the extension of the arctangent function with a principal value in the range of $\left(-\pi,\pi\right]$. The vortex rings modified by Kelvin waves with periodic distortion can be initialised by a three-dimensional wave function
\begin{align}
\psi_{3D}\left(x,y,z\right)=&\psi_{2D}\left\{r-R_1-B_{xy1}{\rm {sin}}[n \theta\left(x,y\right)+n\alpha],\right.\nonumber\\
&\qquad\ \left. z-Z_0-B_{z1}{\rm {cos}}[n \theta(x,y)+n\gamma]\right\}
\times\nonumber\\
&\psi_{2D}^*\left\{r+R_1-B_{xy1}{\rm {sin}}[n \theta(-x,-y)+n\alpha],\right.\nonumber\\
&\qquad\ \left. z-Z_0-B_{z1}{\rm {cos}}[n \theta(-x,-y)+n\gamma]\right\}\times\nonumber\\
&\psi_{2D}\left\{r-R_2-B_{xy2}{\rm {sin}}[n \theta\left(x,y\right)],\right.\nonumber\\
&\qquad\ \left. z+Z_0-B_{z2}\rm {cos}[n \theta(x,y)]\right\}
\times\nonumber\\
&\psi_{2D}^*\left\{r+R_2-B_{xy2}{\rm {sin}}[n \theta(-x,-y)],\right.\nonumber\\
&\qquad\ \left. z+Z_0-B_{z2}{\rm {cos}}[n \theta(-x,-y)]\right\}
\end{align}
where $B_{xyj}$ and $B_{zj}\ (j=1,2)$ are the amplitudes in the radial and axial directions of the Kelvin waves applied for the $j_{th}$ vortex ring, respectively, and $n$ is the wave number of the Kelvin wave perturbations for both rings. The initial rotation of the upper vortex ring around the $z$-axis is given by $\alpha$, and $\gamma$ is the angle of inclination with respect to the unperturbed ring. Then, the minimal offset between the two rings is $d_{min} = 2Z_0-B_{z1}-B_{z2}$. The symmetrical placement of the two coaxial rings with respect to the $xoy$-plane results in $n$ pairs of points between the two rings with the same distance $d_{min}$.

In superfluid systems without fluctuations, the reconnection of those points with a minimum distance occurs simultaneously. In Ref.\,\cite{PNAS.111.15350}, the simultaneous reconnections during the dissolution of knots reduced the complexity of the topology. In Ref.\,\cite{nphys.12.650}, distortions were applied to nots/links, and simultaneous reconnections were avoided during the untying process.

Figures\,\ref{fig:k31}a and b show the relevant parameters for perturbing a vortex ring using helical Kelvin waves, where $R_0$ is the radius of the unperturbed ring, shown in red; $B_{xy}$ and $B_z$ are the amplitudes of the Kelvin perturbations in the $xOy$ plane and $z$ direction, respectively; and $d_{min}$ is the minimal initial offset between the two rings along the propagation axis. If $d_{min}$ is not specially given, we set $d_{min}=a_0$. The relative angle $\alpha$ between the rings is shown in the subplot of Fig.\,\ref{fig:k31}e. Two perturbed rings with radii less than the equilibrium radius\,\cite{PRA.61.013604} and the same winding number move in the same direction. The advantage of this system is that it has a well-controlled initial state that can be created within a spatially confined region, and reconnection occurs naturally due to the different velocities and perturbations of the vortex rings. In addition, the initial parameters can be easily adjusted, which allows us to study the lifetime of the knots/links generated. Figures.\,\ref{fig:k31}c and d show the isosurface and corresponding phase distribution of the trefoil structure in our system as an example.

\section{Results}

\subsection{Generation of a trefoil knot}
To generate a trefoil structure, the rings are initially perturbed by helical Kelvin waves with an odd wave number, $n=3$, co-propagating in the $z$ direction in our system. Simulations are performed with the initial radii of the vortex rings set to $R_1=2a_0$ (blue radius at the top) and $R_2=2.3a_0$ (green radius at the bottom), as shown in Fig.\,\ref{fig:k31}e, and the initial positions of the two rings in the $z$ direction are $z=0.75a_0$ and $z=-0.75a_0$, respectively. We note that the ranges of the radii of the initial vortex rings are limited because small rings cannot resolve the sound pulse and large rings require many grid points\,\cite{PRL.86.1410}. By chosen appropriate parameters, we ensure that the top ring has a higher velocity than the bottom ring, which is an important factor for the association of the two rings. We note that the interaction between rings dominates in our system configuration.

A typical time sequence that illustrates the collision and connections of the initial unlinked vortex rings in Figs.\,\ref{fig:k31}f-h shows some typical topological structures that form during the evolution. As the radii of the rings stretch and shrink during the movement of the rings along the $z$-axis, the top ring catches up with the bottom ring, and connections occur at approximately $t=0.9t_0$ via three simultaneous reconnection events, as shown in Fig.\,\ref{fig:k31}f. A trefoil knot K3-1 with a clear structure and propagation direction identical to that of the initial unlinked rings appears in the condensate cloud, as shown in Fig.\,\ref{fig:k31}g. The trefoil knot exists in the $z<0$ space from $t=1.0t_0$ to $t=3.2t_0$. At approximately $t=3.3t_0$ the knot breaks up via three simultaneous self-reconnection events, as shown in Fig.\,\ref{fig:k31}h, before decaying into two independent rings. In this case, the bridge structures in Figs.\,\ref{fig:k31}f-h are formed through topology-changing reconnections, which alter the topology of the system. In Supplementary Movie 1, we show the time evolution of the vortex structures with another set of initial parameters that yield longer lifetime of the trefoil knot.

\subsection{Generation of a Solomon link}
If the wave number of Kelvin waves is chosen to be an even number, $n=4$, the vortex link L4a1 (Solomon link) is generated via collisions and reconnections of the unlinked vortex rings as shown in Fig.\,\ref{fig:l4a1}. All the other initial conditions are the same as those used in Fig.\,\ref{fig:k31}a. During evolution, the rings perturbed by Kelvin waves distort and reconnect at approximately $t=0.6t_0$, as shown in Fig.\,\ref{fig:l4a1}b. In Fig.\,\ref{fig:l4a1}c, we show that at $t=0.7t_0$, the link L4a1 is generated in the trapped condensate for the first time after the free evolution of the initial rings. However, the link configuration is unstable, and another reconnection event occurs at $t=0.9t_0$, as shown in Fig.\,\ref{fig:l4a1}d. Then, the bridge structure evolves into a vortex link at $t=1.5t_0$, and this link survives much longer than the previous one; another bridge structure appears at $t=2.6t_0$, as shown in Fig.\,\ref{fig:l4a1}f, and then unties to form two highly distorted vortex rings (also see Supplementary Movie 2). We can clearly see that the combination of the vortex rings and the reconnection of the link L4a1 occur through intermediate events with the bridge structures of four simultaneous reconnections, as shown in Figs.\,\ref{fig:l4a1}b, d and f.

\subsection{General cases of knots and links}
In our dynamic process, two types of reconnections exist: topology-changing and topology-conserving reconnections. As shown in Figs.\,\ref{fig:k31}f and h and Figs.\,\ref{fig:l4a1}b and f, the states are topology-changing reconnections, that is, two bridge states for the transitions between different topologies. The state shown in Fig.\,\ref{fig:l4a1}d is classified as a topology-conserving reconnection because the before and after states are all links. The same topology can be obtained with very different geometries. In this paper, we show that for states with the same topology, the geometry is a crucial factor that determines the next step in the evolution. Although the bridge states in Figs.\,\ref{fig:k31}f and h have the same topology and reconnection mechanism, they correspond to inverse dynamical processes.

Notably, the values of the number of the reconnection points, the wave number of the Kelvin waves used to perturb the vortex rings, the crossing number, and the topological writhe of the knot/link generated are the same in our cases. A general rule can thus be discerned: the type of topology generated by two perturbed rings, a knot or a link, is only determined by the parity of the wave number of the Kelvin wave perturbations, which is related to the crossing number and topological writhe of the knots or links generated. Following this rule, we can obtain any type of torus knot or link. Examples with crossing number up to 9, except for 3 and 4, which have already been discussed, are shown in Figs.\,\ref{fig:generalizeresult}a-f, and the types of knots or links obtained are L2a1, K5-1, L6a3, K7-1, L8a14 and K9-1. The standard configuration of torus knots and links are shown in Fig.\,\ref{fig:generalizeresult}g.

\section{Discussion}

\subsection{Stability of knots and links generated}
Knots and links with high crossing number are much more unstable than those with low crossing numbers. As the wave number of helical Kelvin waves increases, the instability of the vortex ring itself increases \cite{PRA.83.045601}, and the decreasing lifetime of the ring, which in turn decreases the lifetimes of the generated knots/links. For $n>6$, the knots/links created decay rapidly. In Ref.\,\cite{science.367.71}, basic topological counting rules were developed
to estimate the relative stability of frequently encountered knots and tangles. In Fig.\,\ref{fig:stability}a, we show the lifetime of the vortex knots/links generated with crossing numbers ranging from 3 to 6. The variation of the lifetime of a trefoil knot with the initial relative angle $\alpha$ between the two vortex rings is shown in Fig.\,\ref{fig:stability}b. Our study shows that there is an ideal angle $\alpha=\pi/n$ for the knots/links with crossing number $n$ that yields a stable state. The blue regions give the time periods when the knot/link survives. The effects of other parameters, such as the initial radius of the bottom ring $R_2$, the planar amplitude of the Kelvin waves $B_{xy2}$, the vertical amplitude of the Kelvin waves $B_{z2}$, and the initial offset $d_{min}$ on the lifetime of the generated knots and links are analysed. The results for the trefoil knot (K3-1) are shown in Fig.\,\ref{fig:stability}b and those for the Solomon link (L4a1) in Fig.\,\ref{fig:stability}c. When studying the effects of one parameter, all the others parameters are held constant. Among these parameters, by choosing an appropriate value of the initial offset $d_{min}$, the topological structure we want can exist much longer, as shown in the last column in Fig.\,\ref{fig:stability}. In a system with a homogeneous background, the intensity of the collision of the collision of vortex rings moving towards each other depends on $d_{min}$\,\cite{PRL.86.1410}. The smaller the offset is, the more violent the collision is, which induces more energy loss during the reconnection process and decreases the energy of the created topological objects. However, there is a threshold for the value of the offset that makes the reconnection of two rings possible.

For the initial radii of the rings, similar values of $R_1$ and $R_2$ make reconnection easier, which results in a longer lifetime for the knots and links. We found that the larger the relative planar distortion of the rings ($B_{xy}$) is, the longer the lifetime of the knots/links is, and the effects of vertical distortion $B_z$ on the knots and links varied.
We note that there should be a set of optimal initial parameters for generating a knot/link with a lifetime as long as possible. The number of events associated with generating or breaking up knots/links also depends on the initial condition chosen. The possibility of links appearing is twice as high as that for knots. However, the lifetimes of the first appearance are all quite short. To ensure that there are no connections between the vortex structures and the condensate surface for all parameters, we limit the time period to $4t_0$. Supplementary Fig.\,S1 and Movie 1 show some long-time dynamics.

\begin{figure*}[htbp]
	\centering
	\includegraphics[width=1.0\textwidth]{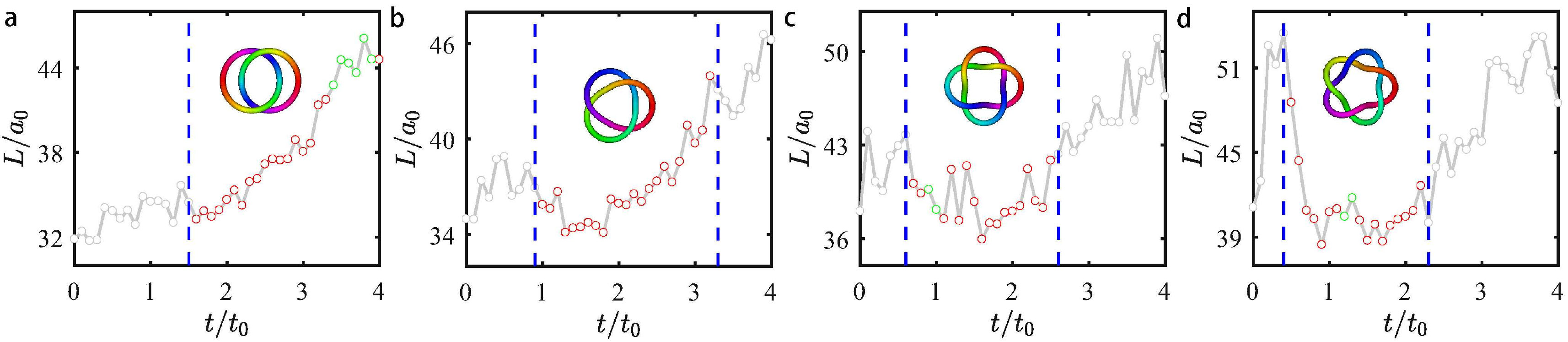}
	\caption{{Length $L$ of the vortex structures as a function of time $t$.} \textbf{a}-\textbf{d}, Unlinking with perturbations by Kelvin waves with wave number $n=2,\ 3,\ 4,\ 5$, respectively. The dashed blue lines indicate the transitions between different topologies. The red open circles indicate the time period when a knot/link (colored patterns) can be clearly identified, and the green circles represent the topology-conserving reconnections.}
	\label{fig:length}
\end{figure*}

\subsection{Length of the vortex structures}

To quantify the vortex dynamics, we compute the vortex length as a function of time as shown in Fig.\,\ref{fig:length}. 
A linked or knotted vortex structure must stretch to expand as it unties in a homogeneous system without boundary\,\cite{nphys.12.650}. In a harmonic trap, when a vortex ring moves from a high density region towards a low density region, the radius of the ring increases gradually\,\cite{PRA.61.013604}. The variation in length reflects the complex interacting process of vortex structures in a harmonic trap. During the untying process of rings/links, the length of these structures increase. However, the decrease in the vortex length during the formation of knots/links is not obvious because this length also increases when these structures move towards the edge of the condensate cloud. Between different topologies, the lengths of vortex structures, except for topology-conserving intermediate structures, vary considerably.

\subsection{Pathways of topological transitions}

In the above discussions, the initial vortex rings are all coaxial, and all the reconnections at different points on the vortex rings occur simultaneously, which results in rings-knot-rings and rings-link-rings transitions. However, the simultaneous reconnections do not affect the intrinsic physics; that is, a transition can develop in both directions between a complex topology and a simple topological in a confined system. In the Supplementary material, we show that by further deforming the initial rings or changing their relative positions, the sequence of reconnection events can be controlled. New pathways of the topology transition are provided, such as rings-link-knot-link-rings and rings-link-ring-link-rings, through successive reconnections, even though the wave number of the Kelvin perturbations $n$ is odd (Supplementary Fig\,S2), which provides further evidence of the transition between a simple topology and a complex topology in both directions.

\section{Conclusion}

In conclusion, we find that torus vortex knots/links can be generated through reconnection of vortex rings in a confined superfluid BEC. Notably, even Kelvin wave numbers produced links and odd Kelvin wave numbers produced knots. In a trapped system, the transition between trivial and complex topologies occured in both directions, while in homogeneous systems without boundaries, vortex knots/links untied monotonically, which simplified their topology at each step\,\cite{PNAS.111.15350,Science.357.487,nphys.9.253}. The trapping potential and Kelvin wave perturbations played synergistic roles in forming and stabilizing vortex knots/links. In supplementary material, we also show that all these results can be achieved in a BCE system with a box trapping potential. The simplicity and robustness of our method provide a new framework for discovering knots/links with complex topologies, which might be of great interest in the strategic design of topologically complex structures with chemical and biological molecules, and in understanding the evolution and mechanisms of confined turbulent systems.

\section*{Acknowledgments}

{We acknowledge funding by the National Science Foundation of China under grants Nos. 11775178, 11947301 and 61835013, by National Key R\&D Program of China under grants No. 2016YFA0301500, the Strategic Priority Research Program of the Chinese Academy of Sciences under grants Nos. XDB01020300 and XDB21030300, the Major Basic Research Program of Natural Science of Shaanxi Province under grants Nos. 2017KCT-12 and 2017ZDJC-32. This research is also supported by the Double First-class University Construction Project of Northwest University. }


\begin{thebibliography}{43}%
\makeatletter
\providecommand \@ifxundefined [1]{%
 \@ifx{#1\undefined}
}%
\providecommand \@ifnum [1]{%
 \ifnum #1\expandafter \@firstoftwo
 \else \expandafter \@secondoftwo
 \fi
}%
\providecommand \@ifx [1]{%
 \ifx #1\expandafter \@firstoftwo
 \else \expandafter \@secondoftwo
 \fi
}%
\providecommand \natexlab [1]{#1}%
\providecommand \enquote  [1]{``#1''}%
\providecommand \bibnamefont  [1]{#1}%
\providecommand \bibfnamefont [1]{#1}%
\providecommand \citenamefont [1]{#1}%
\providecommand \href@noop [0]{\@secondoftwo}%
\providecommand \href [0]{\begingroup \@sanitize@url \@href}%
\providecommand \@href[1]{\@@startlink{#1}\@@href}%
\providecommand \@@href[1]{\endgroup#1\@@endlink}%
\providecommand \@sanitize@url [0]{\catcode `\\12\catcode `\$12\catcode
  `\&12\catcode `\#12\catcode `\^12\catcode `\_12\catcode `\%12\relax}%
\providecommand \@@startlink[1]{}%
\providecommand \@@endlink[0]{}%
\providecommand \url  [0]{\begingroup\@sanitize@url \@url }%
\providecommand \@url [1]{\endgroup\@href {#1}{\urlprefix }}%
\providecommand \urlprefix  [0]{URL }%
\providecommand \Eprint [0]{\href }%
\providecommand \doibase [0]{http://dx.doi.org/}%
\providecommand \selectlanguage [0]{\@gobble}%
\providecommand \bibinfo  [0]{\@secondoftwo}%
\providecommand \bibfield  [0]{\@secondoftwo}%
\providecommand \translation [1]{[#1]}%
\providecommand \BibitemOpen [0]{}%
\providecommand \bibitemStop [0]{}%
\providecommand \bibitemNoStop [0]{.\EOS\space}%
\providecommand \EOS [0]{\spacefactor3000\relax}%
\providecommand \BibitemShut  [1]{\csname bibitem#1\endcsname}%
\let\auto@bib@innerbib\@empty
\bibitem [{\citenamefont {Kleckner}\ \emph {et~al.}(2013)\citenamefont
  {Kleckner}, \citenamefont {Scheeler},\ and\ \citenamefont
  {Irvine}}]{nphys.9.253}%
  \BibitemOpen
  \bibfield  {author} {\bibinfo {author} {\bibfnamefont {D.}~\bibnamefont
  {Kleckner}}, \bibinfo {author} {\bibfnamefont {M.}~\bibnamefont {Scheeler}},
  \ and\ \bibinfo {author} {\bibfnamefont {W.}~\bibnamefont {Irvine}},\ }\href
  {\doibase 10.1038/nphys2560} {\bibfield  {journal} {\bibinfo  {journal}
  {Nature Physics}\ }\textbf {\bibinfo {volume} {9}},\ \bibinfo {pages} {253}
  (\bibinfo {year} {2013})}\BibitemShut {NoStop}%
\bibitem [{\citenamefont {Scheeler}\ \emph {et~al.}(2014)\citenamefont
  {Scheeler}, \citenamefont {Kleckner}, \citenamefont {Proment}, \citenamefont
  {Kindlmann},\ and\ \citenamefont {Irvine}}]{PNAS.111.15350}%
  \BibitemOpen
  \bibfield  {author} {\bibinfo {author} {\bibfnamefont {M.~W.}\ \bibnamefont
  {Scheeler}}, \bibinfo {author} {\bibfnamefont {D.}~\bibnamefont {Kleckner}},
  \bibinfo {author} {\bibfnamefont {D.}~\bibnamefont {Proment}}, \bibinfo
  {author} {\bibfnamefont {G.~L.}\ \bibnamefont {Kindlmann}}, \ and\ \bibinfo
  {author} {\bibfnamefont {W.~T.~M.}\ \bibnamefont {Irvine}},\ }\href {\doibase
  10.1073/pnas.1407232111} {\bibfield  {journal} {\bibinfo  {journal}
  {Proceedings of the National Academy of Sciences}\ }\textbf {\bibinfo
  {volume} {111}},\ \bibinfo {pages} {15350} (\bibinfo {year}
  {2014})}\BibitemShut {NoStop}%
\bibitem [{\citenamefont {Shomroni}\ \emph {et~al.}(2009)\citenamefont
  {Shomroni}, \citenamefont {Lahoud}, \citenamefont {Levy},\ and\ \citenamefont
  {Steinhauer}}]{nphys.5.193}%
  \BibitemOpen
  \bibfield  {author} {\bibinfo {author} {\bibfnamefont {I.}~\bibnamefont
  {Shomroni}}, \bibinfo {author} {\bibfnamefont {E.}~\bibnamefont {Lahoud}},
  \bibinfo {author} {\bibfnamefont {S.}~\bibnamefont {Levy}}, \ and\ \bibinfo
  {author} {\bibfnamefont {J.}~\bibnamefont {Steinhauer}},\ }\href {\doibase
  10.1038/nphys1177} {\bibfield  {journal} {\bibinfo  {journal} {Nature
  Physics}\ }\textbf {\bibinfo {volume} {5}},\ \bibinfo {pages} {193} (\bibinfo
  {year} {2009})}\BibitemShut {NoStop}%
\bibitem [{\citenamefont {Kleckner}\ \emph {et~al.}(2016)\citenamefont
  {Kleckner}, \citenamefont {Kauffman},\ and\ \citenamefont
  {Irvine}}]{nphys.12.650}%
  \BibitemOpen
  \bibfield  {author} {\bibinfo {author} {\bibfnamefont {D.}~\bibnamefont
  {Kleckner}}, \bibinfo {author} {\bibfnamefont {L.~H.}\ \bibnamefont
  {Kauffman}}, \ and\ \bibinfo {author} {\bibfnamefont {W.~T.~M.}\ \bibnamefont
  {Irvine}},\ }\href {\doibase 10.1038/nphys3679} {\bibfield  {journal}
  {\bibinfo  {journal} {Nature Physics}\ }\textbf {\bibinfo {volume} {12}},\
  \bibinfo {pages} {650} (\bibinfo {year} {2016})}\BibitemShut {NoStop}%
\bibitem [{\citenamefont {Hall}\ \emph {et~al.}(2016)\citenamefont {Hall},
  \citenamefont {Ray}, \citenamefont {Tiurev}, \citenamefont {Ruokokoski},
  \citenamefont {Gheorghe},\ and\ \citenamefont {M\"ott\"onen}}]{nphys.12.478}%
  \BibitemOpen
  \bibfield  {author} {\bibinfo {author} {\bibfnamefont {D.~S.}\ \bibnamefont
  {Hall}}, \bibinfo {author} {\bibfnamefont {M.~W.}\ \bibnamefont {Ray}},
  \bibinfo {author} {\bibfnamefont {K.}~\bibnamefont {Tiurev}}, \bibinfo
  {author} {\bibfnamefont {E.}~\bibnamefont {Ruokokoski}}, \bibinfo {author}
  {\bibfnamefont {A.~H.}\ \bibnamefont {Gheorghe}}, \ and\ \bibinfo {author}
  {\bibfnamefont {M.}~\bibnamefont {M\"ott\"onen}},\ }\href {\doibase
  10.1038/nphys3624} {\bibfield  {journal} {\bibinfo  {journal} {Nature
  Physics}\ }\textbf {\bibinfo {volume} {12}},\ \bibinfo {pages} {478}
  (\bibinfo {year} {2016})}\BibitemShut {NoStop}%
\bibitem [{\citenamefont {Maucher}\ \emph {et~al.}(2016)\citenamefont
  {Maucher}, \citenamefont {Gardiner},\ and\ \citenamefont
  {Hughes}}]{NJP.18.063016}%
  \BibitemOpen
  \bibfield  {author} {\bibinfo {author} {\bibfnamefont {F.}~\bibnamefont
  {Maucher}}, \bibinfo {author} {\bibfnamefont {S.~A.}\ \bibnamefont
  {Gardiner}}, \ and\ \bibinfo {author} {\bibfnamefont {I.~G.}\ \bibnamefont
  {Hughes}},\ }\href {\doibase 10.1088/1367-2630/18/6/063016} {\bibfield
  {journal} {\bibinfo  {journal} {New Journal of Physics}\ }\textbf {\bibinfo
  {volume} {18}},\ \bibinfo {pages} {063016} (\bibinfo {year}
  {2016})}\BibitemShut {NoStop}%
\bibitem [{\citenamefont {Proment}\ \emph {et~al.}(2012)\citenamefont
  {Proment}, \citenamefont {Onorato},\ and\ \citenamefont
  {Barenghi}}]{PRE.85.036306}%
  \BibitemOpen
  \bibfield  {author} {\bibinfo {author} {\bibfnamefont {D.}~\bibnamefont
  {Proment}}, \bibinfo {author} {\bibfnamefont {M.}~\bibnamefont {Onorato}}, \
  and\ \bibinfo {author} {\bibfnamefont {C.~F.}\ \bibnamefont {Barenghi}},\
  }\href {\doibase 10.1103/PhysRevE.85.036306} {\bibfield  {journal} {\bibinfo
  {journal} {Phys. Rev. E}\ }\textbf {\bibinfo {volume} {85}},\ \bibinfo
  {pages} {036306} (\bibinfo {year} {2012})}\BibitemShut {NoStop}%
\bibitem [{\citenamefont {Moffatt}(2014)}]{PNAS.111.3663}%
  \BibitemOpen
  \bibfield  {author} {\bibinfo {author} {\bibfnamefont {H.~K.}\ \bibnamefont
  {Moffatt}},\ }\href {\doibase 10.1073/pnas.1400277111} {\bibfield  {journal}
  {\bibinfo  {journal} {Proceedings of the National Academy of Sciences}\
  }\textbf {\bibinfo {volume} {111}},\ \bibinfo {pages} {3663} (\bibinfo {year}
  {2014})}\BibitemShut {NoStop}%
\bibitem [{\citenamefont {Ricca}\ and\ \citenamefont
  {Berger}(1996)}]{phys.tod.49.28}%
  \BibitemOpen
  \bibfield  {author} {\bibinfo {author} {\bibfnamefont {R.~L.}\ \bibnamefont
  {Ricca}}\ and\ \bibinfo {author} {\bibfnamefont {M.~A.}\ \bibnamefont
  {Berger}},\ }\href {\doibase 10.1063/1.881574} {\bibfield  {journal}
  {\bibinfo  {journal} {Physics Today}\ }\textbf {\bibinfo {volume} {49}},\
  \bibinfo {pages} {28} (\bibinfo {year} {1996})}\BibitemShut {NoStop}%
\bibitem [{\citenamefont {Raymer}\ and\ \citenamefont
  {Smith}(2007)}]{PNAS.104.16432}%
  \BibitemOpen
  \bibfield  {author} {\bibinfo {author} {\bibfnamefont {D.~M.}\ \bibnamefont
  {Raymer}}\ and\ \bibinfo {author} {\bibfnamefont {D.~E.}\ \bibnamefont
  {Smith}},\ }\href {\doibase 10.1073/pnas.0611320104} {\bibfield  {journal}
  {\bibinfo  {journal} {Proceedings of the National Academy of Sciences}\
  }\textbf {\bibinfo {volume} {104}},\ \bibinfo {pages} {16432} (\bibinfo
  {year} {2007})}\BibitemShut {NoStop}%
\bibitem [{\citenamefont {Tkalec}\ \emph {et~al.}(2011)\citenamefont {Tkalec},
  \citenamefont {Ravnik}, \citenamefont {{\v C}opar}, \citenamefont {{\v
  Z}umer},\ and\ \citenamefont {Mu{\v s}evi{\v c}}}]{Science.333.62}%
  \BibitemOpen
  \bibfield  {author} {\bibinfo {author} {\bibfnamefont {U.}~\bibnamefont
  {Tkalec}}, \bibinfo {author} {\bibfnamefont {M.}~\bibnamefont {Ravnik}},
  \bibinfo {author} {\bibfnamefont {S.}~\bibnamefont {{\v C}opar}}, \bibinfo
  {author} {\bibfnamefont {S.}~\bibnamefont {{\v Z}umer}}, \ and\ \bibinfo
  {author} {\bibfnamefont {I.}~\bibnamefont {Mu{\v s}evi{\v c}}},\ }\href
  {\doibase 10.1126/science.1205705} {\bibfield  {journal} {\bibinfo  {journal}
  {Science}\ }\textbf {\bibinfo {volume} {333}},\ \bibinfo {pages} {62}
  (\bibinfo {year} {2011})}\BibitemShut {NoStop}%
\bibitem [{\citenamefont {Martinez}\ \emph {et~al.}(2014)\citenamefont
  {Martinez}, \citenamefont {Ravnik}, \citenamefont {Lucero}, \citenamefont
  {Visvanathan}, \citenamefont {{\v Z}umer},\ and\ \citenamefont
  {Smalyukh}}]{nmat.13.258}%
  \BibitemOpen
  \bibfield  {author} {\bibinfo {author} {\bibfnamefont {A.}~\bibnamefont
  {Martinez}}, \bibinfo {author} {\bibfnamefont {M.}~\bibnamefont {Ravnik}},
  \bibinfo {author} {\bibfnamefont {B.}~\bibnamefont {Lucero}}, \bibinfo
  {author} {\bibfnamefont {R.}~\bibnamefont {Visvanathan}}, \bibinfo {author}
  {\bibfnamefont {S.}~\bibnamefont {{\v Z}umer}}, \ and\ \bibinfo {author}
  {\bibfnamefont {I.~I.}\ \bibnamefont {Smalyukh}},\ }\href {\doibase
  10.1038/nmat3840} {\bibfield  {journal} {\bibinfo  {journal} {Nature
  Materials}\ }\textbf {\bibinfo {volume} {13}},\ \bibinfo {pages} {258}
  (\bibinfo {year} {2014})}\BibitemShut {NoStop}%
\bibitem [{\citenamefont {Wasserman}\ and\ \citenamefont
  {Cozzarelli}(1986)}]{Science.232.951}%
  \BibitemOpen
  \bibfield  {author} {\bibinfo {author} {\bibfnamefont {S.}~\bibnamefont
  {Wasserman}}\ and\ \bibinfo {author} {\bibfnamefont {N.}~\bibnamefont
  {Cozzarelli}},\ }\href {\doibase 10.1126/science.3010458} {\bibfield
  {journal} {\bibinfo  {journal} {Science}\ }\textbf {\bibinfo {volume}
  {232}},\ \bibinfo {pages} {951} (\bibinfo {year} {1986})}\BibitemShut
  {NoStop}%
\bibitem [{\citenamefont {Schlick}\ and\ \citenamefont
  {Olson}(1992)}]{science.257.1110}%
  \BibitemOpen
  \bibfield  {author} {\bibinfo {author} {\bibfnamefont {T.}~\bibnamefont
  {Schlick}}\ and\ \bibinfo {author} {\bibfnamefont {W.~K.}\ \bibnamefont
  {Olson}},\ }\href {\doibase 10.1126/science.257.5073.1110} {\bibfield
  {journal} {\bibinfo  {journal} {Science}\ }\textbf {\bibinfo {volume}
  {257}},\ \bibinfo {pages} {1110} (\bibinfo {year} {1992})}\BibitemShut
  {NoStop}%
\bibitem [{\citenamefont {Dongran}\ \emph {et~al.}(2010)\citenamefont
  {Dongran}, \citenamefont {Suchetan}, \citenamefont {Yan},\ and\ \citenamefont
  {Hao}}]{nnano.5.712}%
  \BibitemOpen
  \bibfield  {author} {\bibinfo {author} {\bibfnamefont {H.}~\bibnamefont
  {Dongran}}, \bibinfo {author} {\bibfnamefont {P.}~\bibnamefont {Suchetan}},
  \bibinfo {author} {\bibfnamefont {L.}~\bibnamefont {Yan}}, \ and\ \bibinfo
  {author} {\bibfnamefont {Y.}~\bibnamefont {Hao}},\ }\href
  {https://doi.org/10.1038/nnano.2010.193} {\bibfield  {journal} {\bibinfo
  {journal} {Nature Nanotechnology}\ }\textbf {\bibinfo {volume} {5}},\
  \bibinfo {pages} {712} (\bibinfo {year} {2010})}\BibitemShut {NoStop}%
\bibitem [{\citenamefont {Chichak}\ \emph {et~al.}(2004)\citenamefont
  {Chichak}, \citenamefont {Cantrill}, \citenamefont {Pease}, \citenamefont
  {Chiu}, \citenamefont {Cave}, \citenamefont {Atwood},\ and\ \citenamefont
  {Stoddart}}]{Science.304.1308}%
  \BibitemOpen
  \bibfield  {author} {\bibinfo {author} {\bibfnamefont {K.~S.}\ \bibnamefont
  {Chichak}}, \bibinfo {author} {\bibfnamefont {S.~J.}\ \bibnamefont
  {Cantrill}}, \bibinfo {author} {\bibfnamefont {A.~R.}\ \bibnamefont {Pease}},
  \bibinfo {author} {\bibfnamefont {S.-H.}\ \bibnamefont {Chiu}}, \bibinfo
  {author} {\bibfnamefont {G.~W.~V.}\ \bibnamefont {Cave}}, \bibinfo {author}
  {\bibfnamefont {J.~L.}\ \bibnamefont {Atwood}}, \ and\ \bibinfo {author}
  {\bibfnamefont {J.~F.}\ \bibnamefont {Stoddart}},\ }\href {\doibase
  10.1126/science.1096914} {\bibfield  {journal} {\bibinfo  {journal}
  {Science}\ }\textbf {\bibinfo {volume} {304}},\ \bibinfo {pages} {1308}
  (\bibinfo {year} {2004})}\BibitemShut {NoStop}%
\bibitem [{\citenamefont {Sumners}(1995)}]{NAMS.42.528}%
  \BibitemOpen
  \bibfield  {author} {\bibinfo {author} {\bibfnamefont {D.~W.}\ \bibnamefont
  {Sumners}},\ }\href@noop {} {\bibfield  {journal} {\bibinfo  {journal}
  {Notices of the AMS}\ }\textbf {\bibinfo {volume} {42}},\ \bibinfo {pages}
  {528} (\bibinfo {year} {1995})}\BibitemShut {NoStop}%
\bibitem [{\citenamefont {Shimokawa}\ \emph {et~al.}(2013)\citenamefont
  {Shimokawa}, \citenamefont {Ishihara}, \citenamefont {Grainge}, \citenamefont
  {Sherratt},\ and\ \citenamefont {Vazquez}}]{PNAS.110.20906}%
  \BibitemOpen
  \bibfield  {author} {\bibinfo {author} {\bibfnamefont {K.}~\bibnamefont
  {Shimokawa}}, \bibinfo {author} {\bibfnamefont {K.}~\bibnamefont {Ishihara}},
  \bibinfo {author} {\bibfnamefont {I.}~\bibnamefont {Grainge}}, \bibinfo
  {author} {\bibfnamefont {D.~J.}\ \bibnamefont {Sherratt}}, \ and\ \bibinfo
  {author} {\bibfnamefont {M.}~\bibnamefont {Vazquez}},\ }\href {\doibase
  10.1073/pnas.1308450110} {\bibfield  {journal} {\bibinfo  {journal}
  {Proceedings of the National Academy of Sciences}\ }\textbf {\bibinfo
  {volume} {110}},\ \bibinfo {pages} {20906} (\bibinfo {year}
  {2013})}\BibitemShut {NoStop}%
\bibitem [{\citenamefont {Liu}\ \emph {et~al.}(2017)\citenamefont {Liu},
  \citenamefont {Shao}, \citenamefont {Chen}, \citenamefont {Tse-Dinh},
  \citenamefont {Piccirilli},\ and\ \citenamefont {Weizmann}}]{ncomm.8.14936}%
  \BibitemOpen
  \bibfield  {author} {\bibinfo {author} {\bibfnamefont {D.}~\bibnamefont
  {Liu}}, \bibinfo {author} {\bibfnamefont {Y.}~\bibnamefont {Shao}}, \bibinfo
  {author} {\bibfnamefont {G.}~\bibnamefont {Chen}}, \bibinfo {author}
  {\bibfnamefont {Y.-C.}\ \bibnamefont {Tse-Dinh}}, \bibinfo {author}
  {\bibfnamefont {J.~A.}\ \bibnamefont {Piccirilli}}, \ and\ \bibinfo {author}
  {\bibfnamefont {Y.}~\bibnamefont {Weizmann}},\ }\href {\doibase
  10.1038/ncomms14936} {\bibfield  {journal} {\bibinfo  {journal} {Nature
  Communications}\ }\textbf {\bibinfo {volume} {8}},\ \bibinfo {pages} {14936}
  (\bibinfo {year} {2017})}\BibitemShut {NoStop}%
\bibitem [{\citenamefont {Forgan}\ \emph {et~al.}(2011)\citenamefont {Forgan},
  \citenamefont {Sauvage},\ and\ \citenamefont {Stoddart}}]{Chem.Rev.111.5434}%
  \BibitemOpen
  \bibfield  {author} {\bibinfo {author} {\bibfnamefont {R.~S.}\ \bibnamefont
  {Forgan}}, \bibinfo {author} {\bibfnamefont {J.-P.}\ \bibnamefont {Sauvage}},
  \ and\ \bibinfo {author} {\bibfnamefont {J.~F.}\ \bibnamefont {Stoddart}},\
  }\href {\doibase 10.1021/cr200034u} {\bibfield  {journal} {\bibinfo
  {journal} {Chemical Reviews}\ }\textbf {\bibinfo {volume} {111}},\ \bibinfo
  {pages} {5434} (\bibinfo {year} {2011})}\BibitemShut {NoStop}%
\bibitem [{\citenamefont {Marcos}\ \emph {et~al.}(2016)\citenamefont {Marcos},
  \citenamefont {Stephens}, \citenamefont {Jaramillo-Garcia}, \citenamefont
  {Nussbaumer}, \citenamefont {Woltering}, \citenamefont {Valero},
  \citenamefont {Lemonnier}, \citenamefont {Vitorica-Yrezabal},\ and\
  \citenamefont {Leigh}}]{science.352.1555}%
  \BibitemOpen
  \bibfield  {author} {\bibinfo {author} {\bibfnamefont {V.}~\bibnamefont
  {Marcos}}, \bibinfo {author} {\bibfnamefont {A.~J.}\ \bibnamefont
  {Stephens}}, \bibinfo {author} {\bibfnamefont {J.}~\bibnamefont
  {Jaramillo-Garcia}}, \bibinfo {author} {\bibfnamefont {A.~L.}\ \bibnamefont
  {Nussbaumer}}, \bibinfo {author} {\bibfnamefont {S.~L.}\ \bibnamefont
  {Woltering}}, \bibinfo {author} {\bibfnamefont {A.}~\bibnamefont {Valero}},
  \bibinfo {author} {\bibfnamefont {J.-F.}\ \bibnamefont {Lemonnier}}, \bibinfo
  {author} {\bibfnamefont {I.~J.}\ \bibnamefont {Vitorica-Yrezabal}}, \ and\
  \bibinfo {author} {\bibfnamefont {D.~A.}\ \bibnamefont {Leigh}},\ }\href
  {\doibase 10.1126/science.aaf3673} {\bibfield  {journal} {\bibinfo  {journal}
  {Science}\ }\textbf {\bibinfo {volume} {352}},\ \bibinfo {pages} {1555}
  (\bibinfo {year} {2016})}\BibitemShut {NoStop}%
\bibitem [{\citenamefont {Danon}\ \emph {et~al.}(2017)\citenamefont {Danon},
  \citenamefont {Kr{\"u}ger}, \citenamefont {Leigh}, \citenamefont {Lemonnier},
  \citenamefont {Stephens}, \citenamefont {Vitorica-Yrezabal},\ and\
  \citenamefont {Woltering}}]{science.355.159}%
  \BibitemOpen
  \bibfield  {author} {\bibinfo {author} {\bibfnamefont {J.~J.}\ \bibnamefont
  {Danon}}, \bibinfo {author} {\bibfnamefont {A.}~\bibnamefont {Kr{\"u}ger}},
  \bibinfo {author} {\bibfnamefont {D.~A.}\ \bibnamefont {Leigh}}, \bibinfo
  {author} {\bibfnamefont {J.-F.}\ \bibnamefont {Lemonnier}}, \bibinfo {author}
  {\bibfnamefont {A.~J.}\ \bibnamefont {Stephens}}, \bibinfo {author}
  {\bibfnamefont {I.~J.}\ \bibnamefont {Vitorica-Yrezabal}}, \ and\ \bibinfo
  {author} {\bibfnamefont {S.~L.}\ \bibnamefont {Woltering}},\ }\href {\doibase
  10.1126/science.aal1619} {\bibfield  {journal} {\bibinfo  {journal}
  {Science}\ }\textbf {\bibinfo {volume} {355}},\ \bibinfo {pages} {159}
  (\bibinfo {year} {2017})}\BibitemShut {NoStop}%
\bibitem [{\citenamefont {Segawa}\ \emph {et~al.}(2019)\citenamefont {Segawa},
  \citenamefont {Kuwayama}, \citenamefont {Hijikata}, \citenamefont {Fushimi},
  \citenamefont {Nishihara}, \citenamefont {Pirillo}, \citenamefont
  {Shirasaki}, \citenamefont {Kubota},\ and\ \citenamefont
  {Itami}}]{Science.365.272}%
  \BibitemOpen
  \bibfield  {author} {\bibinfo {author} {\bibfnamefont {Y.}~\bibnamefont
  {Segawa}}, \bibinfo {author} {\bibfnamefont {M.}~\bibnamefont {Kuwayama}},
  \bibinfo {author} {\bibfnamefont {Y.}~\bibnamefont {Hijikata}}, \bibinfo
  {author} {\bibfnamefont {M.}~\bibnamefont {Fushimi}}, \bibinfo {author}
  {\bibfnamefont {T.}~\bibnamefont {Nishihara}}, \bibinfo {author}
  {\bibfnamefont {J.}~\bibnamefont {Pirillo}}, \bibinfo {author} {\bibfnamefont
  {J.}~\bibnamefont {Shirasaki}}, \bibinfo {author} {\bibfnamefont
  {N.}~\bibnamefont {Kubota}}, \ and\ \bibinfo {author} {\bibfnamefont
  {K.}~\bibnamefont {Itami}},\ }\href {\doibase 10.1126/science.aav5021}
  {\bibfield  {journal} {\bibinfo  {journal} {Science}\ }\textbf {\bibinfo
  {volume} {365}},\ \bibinfo {pages} {272} (\bibinfo {year}
  {2019})}\BibitemShut {NoStop}%
\bibitem [{\citenamefont {Scheeler}\ \emph {et~al.}(2017)\citenamefont
  {Scheeler}, \citenamefont {van Rees}, \citenamefont {Kedia}, \citenamefont
  {Kleckner},\ and\ \citenamefont {Irvine}}]{Science.357.487}%
  \BibitemOpen
  \bibfield  {author} {\bibinfo {author} {\bibfnamefont {M.~W.}\ \bibnamefont
  {Scheeler}}, \bibinfo {author} {\bibfnamefont {W.~M.}\ \bibnamefont {van
  Rees}}, \bibinfo {author} {\bibfnamefont {H.}~\bibnamefont {Kedia}}, \bibinfo
  {author} {\bibfnamefont {D.}~\bibnamefont {Kleckner}}, \ and\ \bibinfo
  {author} {\bibfnamefont {W.~T.~M.}\ \bibnamefont {Irvine}},\ }\href {\doibase
  10.1126/science.aam6897} {\bibfield  {journal} {\bibinfo  {journal}
  {Science}\ }\textbf {\bibinfo {volume} {357}},\ \bibinfo {pages} {487}
  (\bibinfo {year} {2017})}\BibitemShut {NoStop}%
\bibitem [{\citenamefont {Dennis}\ \emph {et~al.}(2010)\citenamefont {Dennis},
  \citenamefont {King}, \citenamefont {Jack}, \citenamefont {O'Holleran},\ and\
  \citenamefont {Padgett}}]{nphys.6.118}%
  \BibitemOpen
  \bibfield  {author} {\bibinfo {author} {\bibfnamefont {M.~R.}\ \bibnamefont
  {Dennis}}, \bibinfo {author} {\bibfnamefont {R.~P.}\ \bibnamefont {King}},
  \bibinfo {author} {\bibfnamefont {B.}~\bibnamefont {Jack}}, \bibinfo {author}
  {\bibfnamefont {K.}~\bibnamefont {O'Holleran}}, \ and\ \bibinfo {author}
  {\bibfnamefont {M.~J.}\ \bibnamefont {Padgett}},\ }\href {\doibase
  10.1038/nphys1504} {\bibfield  {journal} {\bibinfo  {journal} {Nature
  Physics}\ }\textbf {\bibinfo {volume} {6}},\ \bibinfo {pages} {118} (\bibinfo
  {year} {2010})}\BibitemShut {NoStop}%
\bibitem [{\citenamefont {Liu}\ and\ \citenamefont
  {Ricca}(2016)}]{srep.6.24118}%
  \BibitemOpen
  \bibfield  {author} {\bibinfo {author} {\bibfnamefont {X.}~\bibnamefont
  {Liu}}\ and\ \bibinfo {author} {\bibfnamefont {R.~L.}\ \bibnamefont
  {Ricca}},\ }\href {\doibase 10.1038/srep24118} {\bibfield  {journal}
  {\bibinfo  {journal} {Scientific Reports}\ }\textbf {\bibinfo {volume} {6}},\
  \bibinfo {pages} {24118} (\bibinfo {year} {2016})}\BibitemShut {NoStop}%
\bibitem [{\citenamefont {Stolz}\ \emph {et~al.}(2017)\citenamefont {Stolz},
  \citenamefont {Yoshida}, \citenamefont {Brasher}, \citenamefont {Flanner},
  \citenamefont {Ishihara}, \citenamefont {Sherratt}, \citenamefont
  {Shimokawa},\ and\ \citenamefont {Vazquez}}]{srep.7.12420}%
  \BibitemOpen
  \bibfield  {author} {\bibinfo {author} {\bibfnamefont {R.}~\bibnamefont
  {Stolz}}, \bibinfo {author} {\bibfnamefont {M.}~\bibnamefont {Yoshida}},
  \bibinfo {author} {\bibfnamefont {R.}~\bibnamefont {Brasher}}, \bibinfo
  {author} {\bibfnamefont {M.}~\bibnamefont {Flanner}, \bibfnamefont
  {MichelleFlanner}}, \bibinfo {author} {\bibfnamefont {K.}~\bibnamefont
  {Ishihara}}, \bibinfo {author} {\bibfnamefont {D.~J.}\ \bibnamefont
  {Sherratt}}, \bibinfo {author} {\bibfnamefont {K.}~\bibnamefont {Shimokawa}},
  \ and\ \bibinfo {author} {\bibfnamefont {M.}~\bibnamefont {Vazquez}},\ }\href
  {\doibase 10.1038/s41598-017-12172-2} {\bibfield  {journal} {\bibinfo
  {journal} {Scientific Reports}\ }\textbf {\bibinfo {volume} {7}},\ \bibinfo
  {pages} {12420} (\bibinfo {year} {2017})}\BibitemShut {NoStop}%
\bibitem [{\citenamefont {Zuccher}\ and\ \citenamefont
  {Ricca}(2017)}]{PRE.95.053109}%
  \BibitemOpen
  \bibfield  {author} {\bibinfo {author} {\bibfnamefont {S.}~\bibnamefont
  {Zuccher}}\ and\ \bibinfo {author} {\bibfnamefont {R.~L.}\ \bibnamefont
  {Ricca}},\ }\href {\doibase 10.1103/PhysRevE.95.053109} {\bibfield  {journal}
  {\bibinfo  {journal} {Phys. Rev. E}\ }\textbf {\bibinfo {volume} {95}},\
  \bibinfo {pages} {053109} (\bibinfo {year} {2017})}\BibitemShut {NoStop}%
\bibitem [{\citenamefont {Lim}\ and\ \citenamefont
  {Nickels}(1992)}]{nature.357.225}%
  \BibitemOpen
  \bibfield  {author} {\bibinfo {author} {\bibfnamefont {T.~T.}\ \bibnamefont
  {Lim}}\ and\ \bibinfo {author} {\bibfnamefont {T.~B.}\ \bibnamefont
  {Nickels}},\ }\href {\doibase 10.1038/357225a0} {\bibfield  {journal}
  {\bibinfo  {journal} {Nature}\ }\textbf {\bibinfo {volume} {357}},\ \bibinfo
  {pages} {2225} (\bibinfo {year} {1992})}\BibitemShut {NoStop}%
\bibitem [{\citenamefont {Koplik}\ and\ \citenamefont
  {Levine}(1996)}]{PRL.76.4745}%
  \BibitemOpen
  \bibfield  {author} {\bibinfo {author} {\bibfnamefont {J.}~\bibnamefont
  {Koplik}}\ and\ \bibinfo {author} {\bibfnamefont {H.}~\bibnamefont
  {Levine}},\ }\href {\doibase 10.1103/PhysRevLett.76.4745} {\bibfield
  {journal} {\bibinfo  {journal} {Phys. Rev. Lett.}\ }\textbf {\bibinfo
  {volume} {76}},\ \bibinfo {pages} {4745} (\bibinfo {year}
  {1996})}\BibitemShut {NoStop}%
\bibitem [{\citenamefont {Serafini}\ \emph {et~al.}(2017)\citenamefont
  {Serafini}, \citenamefont {Galantucci}, \citenamefont {Iseni}, \citenamefont
  {Bienaim\'e}, \citenamefont {Bisset}, \citenamefont {Barenghi}, \citenamefont
  {Dalfovo}, \citenamefont {Lamporesi},\ and\ \citenamefont
  {Ferrari}}]{PRX.7.021031}%
  \BibitemOpen
  \bibfield  {author} {\bibinfo {author} {\bibfnamefont {S.}~\bibnamefont
  {Serafini}}, \bibinfo {author} {\bibfnamefont {L.}~\bibnamefont
  {Galantucci}}, \bibinfo {author} {\bibfnamefont {E.}~\bibnamefont {Iseni}},
  \bibinfo {author} {\bibfnamefont {T.}~\bibnamefont {Bienaim\'e}}, \bibinfo
  {author} {\bibfnamefont {R.~N.}\ \bibnamefont {Bisset}}, \bibinfo {author}
  {\bibfnamefont {C.~F.}\ \bibnamefont {Barenghi}}, \bibinfo {author}
  {\bibfnamefont {F.}~\bibnamefont {Dalfovo}}, \bibinfo {author} {\bibfnamefont
  {G.}~\bibnamefont {Lamporesi}}, \ and\ \bibinfo {author} {\bibfnamefont
  {G.}~\bibnamefont {Ferrari}},\ }\href {\doibase 10.1103/PhysRevX.7.021031}
  {\bibfield  {journal} {\bibinfo  {journal} {Phys. Rev. X}\ }\textbf {\bibinfo
  {volume} {7}},\ \bibinfo {pages} {021031} (\bibinfo {year}
  {2017})}\BibitemShut {NoStop}%
\bibitem [{\citenamefont {Kivotides}\ \emph {et~al.}(2001)\citenamefont
  {Kivotides}, \citenamefont {Vassilicos}, \citenamefont {Samuels},\ and\
  \citenamefont {Barenghi}}]{PRL.86.3080}%
  \BibitemOpen
  \bibfield  {author} {\bibinfo {author} {\bibfnamefont {D.}~\bibnamefont
  {Kivotides}}, \bibinfo {author} {\bibfnamefont {J.~C.}\ \bibnamefont
  {Vassilicos}}, \bibinfo {author} {\bibfnamefont {D.~C.}\ \bibnamefont
  {Samuels}}, \ and\ \bibinfo {author} {\bibfnamefont {C.~F.}\ \bibnamefont
  {Barenghi}},\ }\href {\doibase 10.1103/PhysRevLett.86.3080} {\bibfield
  {journal} {\bibinfo  {journal} {Phys. Rev. Lett.}\ }\textbf {\bibinfo
  {volume} {86}},\ \bibinfo {pages} {3080} (\bibinfo {year}
  {2001})}\BibitemShut {NoStop}%
\bibitem [{\citenamefont {Fetter}\ and\ \citenamefont
  {Svidzinsky}(2001)}]{JPCM.13.R135}%
  \BibitemOpen
  \bibfield  {author} {\bibinfo {author} {\bibfnamefont {A.~L.}\ \bibnamefont
  {Fetter}}\ and\ \bibinfo {author} {\bibfnamefont {A.~A.}\ \bibnamefont
  {Svidzinsky}},\ }\href {http://stacks.iop.org/0953-8984/13/i=12/a=201}
  {\bibfield  {journal} {\bibinfo  {journal} {Journal of Physics: Condensed
  Matter}\ }\textbf {\bibinfo {volume} {13}},\ \bibinfo {pages} {R135}
  (\bibinfo {year} {2001})}\BibitemShut {NoStop}%
\bibitem [{\citenamefont {Barenghi}\ \emph {et~al.}(2006)\citenamefont
  {Barenghi}, \citenamefont {H\"anninen},\ and\ \citenamefont
  {Tsubota}}]{PRE.74.046303}%
  \BibitemOpen
  \bibfield  {author} {\bibinfo {author} {\bibfnamefont {C.~F.}\ \bibnamefont
  {Barenghi}}, \bibinfo {author} {\bibfnamefont {R.}~\bibnamefont
  {H\"anninen}}, \ and\ \bibinfo {author} {\bibfnamefont {M.}~\bibnamefont
  {Tsubota}},\ }\href {\doibase 10.1103/PhysRevE.74.046303} {\bibfield
  {journal} {\bibinfo  {journal} {Phys. Rev. E}\ }\textbf {\bibinfo {volume}
  {74}},\ \bibinfo {pages} {046303} (\bibinfo {year} {2006})}\BibitemShut
  {NoStop}%
\bibitem [{\citenamefont {Helm}\ \emph {et~al.}(2011)\citenamefont {Helm},
  \citenamefont {Barenghi},\ and\ \citenamefont {Youd}}]{PRA.83.045601}%
  \BibitemOpen
  \bibfield  {author} {\bibinfo {author} {\bibfnamefont {J.~L.}\ \bibnamefont
  {Helm}}, \bibinfo {author} {\bibfnamefont {C.~F.}\ \bibnamefont {Barenghi}},
  \ and\ \bibinfo {author} {\bibfnamefont {A.~J.}\ \bibnamefont {Youd}},\
  }\href {\doibase 10.1103/PhysRevA.83.045601} {\bibfield  {journal} {\bibinfo
  {journal} {Phys. Rev. A}\ }\textbf {\bibinfo {volume} {83}},\ \bibinfo
  {pages} {045601} (\bibinfo {year} {2011})}\BibitemShut {NoStop}%
\bibitem [{\citenamefont {Moffatt}(1969)}]{JFM.35.117}%
  \BibitemOpen
  \bibfield  {author} {\bibinfo {author} {\bibfnamefont {H.~K.}\ \bibnamefont
  {Moffatt}},\ }\href {\doibase 10.1017/S0022112069000991} {\bibfield
  {journal} {\bibinfo  {journal} {Journal of Fluid Mechanics}\ }\textbf
  {\bibinfo {volume} {35}},\ \bibinfo {pages} {117} (\bibinfo {year}
  {1969})}\BibitemShut {NoStop}%
\bibitem [{\citenamefont {Niemi}(2005)}]{PRL.94.124502}%
  \BibitemOpen
  \bibfield  {author} {\bibinfo {author} {\bibfnamefont {A.~J.}\ \bibnamefont
  {Niemi}},\ }\href {\doibase 10.1103/PhysRevLett.94.124502} {\bibfield
  {journal} {\bibinfo  {journal} {Phys. Rev. Lett.}\ }\textbf {\bibinfo
  {volume} {94}},\ \bibinfo {pages} {124502} (\bibinfo {year}
  {2005})}\BibitemShut {NoStop}%
\bibitem [{\citenamefont {Laing}\ \emph {et~al.}(2015)\citenamefont {Laing},
  \citenamefont {Ricca},\ and\ \citenamefont {Sumners}}]{srep.5.9224}%
  \BibitemOpen
  \bibfield  {author} {\bibinfo {author} {\bibfnamefont {C.~E.}\ \bibnamefont
  {Laing}}, \bibinfo {author} {\bibfnamefont {R.~L.}\ \bibnamefont {Ricca}}, \
  and\ \bibinfo {author} {\bibfnamefont {D.~W.~L.}\ \bibnamefont {Sumners}},\
  }\href {\doibase 10.1038/srep09224} {\bibfield  {journal} {\bibinfo
  {journal} {Scientific Reports}\ }\textbf {\bibinfo {volume} {5}},\ \bibinfo
  {pages} {9224} (\bibinfo {year} {2015})}\BibitemShut {NoStop}%
\bibitem [{\citenamefont {Vazquez}\ and\ \citenamefont
  {Summers}(2004)}]{Maths.Proc.Camb.Phil.Soc.136.565}%
  \BibitemOpen
  \bibfield  {author} {\bibinfo {author} {\bibfnamefont {M.}~\bibnamefont
  {Vazquez}}\ and\ \bibinfo {author} {\bibfnamefont {D.~W.}\ \bibnamefont
  {Summers}},\ }\href {\doibase 10.1017/S0305004103007266} {\bibfield
  {journal} {\bibinfo  {journal} {Math. Proc. Camb. Phil. Soc.}\ }\textbf
  {\bibinfo {volume} {136}},\ \bibinfo {pages} {565} (\bibinfo {year}
  {2004})}\BibitemShut {NoStop}%
\bibitem [{\citenamefont {Galantucci}\ \emph {et~al.}(2019)\citenamefont
  {Galantucci}, \citenamefont {Baggaley}, \citenamefont {Parker},\ and\
  \citenamefont {Barenghi}}]{PNAS.116.12204}%
  \BibitemOpen
  \bibfield  {author} {\bibinfo {author} {\bibfnamefont {L.}~\bibnamefont
  {Galantucci}}, \bibinfo {author} {\bibfnamefont {A.~W.}\ \bibnamefont
  {Baggaley}}, \bibinfo {author} {\bibfnamefont {N.~G.}\ \bibnamefont
  {Parker}}, \ and\ \bibinfo {author} {\bibfnamefont {C.~F.}\ \bibnamefont
  {Barenghi}},\ }\href {\doibase 10.1073/pnas.1818668116} {\bibfield  {journal}
  {\bibinfo  {journal} {Proceedings of the National Academy of Sciences}\
  }\textbf {\bibinfo {volume} {116}},\ \bibinfo {pages} {12204} (\bibinfo
  {year} {2019})}\BibitemShut {NoStop}%
\bibitem [{\citenamefont {Jackson}\ \emph {et~al.}(1999)\citenamefont
  {Jackson}, \citenamefont {McCann},\ and\ \citenamefont
  {Adams}}]{PRA.61.013604}%
  \BibitemOpen
  \bibfield  {author} {\bibinfo {author} {\bibfnamefont {B.}~\bibnamefont
  {Jackson}}, \bibinfo {author} {\bibfnamefont {J.~F.}\ \bibnamefont {McCann}},
  \ and\ \bibinfo {author} {\bibfnamefont {C.~S.}\ \bibnamefont {Adams}},\
  }\href {\doibase 10.1103/PRA.61.013604} {\bibfield  {journal} {\bibinfo
  {journal} {Phys. Rev. A}\ }\textbf {\bibinfo {volume} {61}},\ \bibinfo
  {pages} {013604} (\bibinfo {year} {1999})}\BibitemShut {NoStop}%
\bibitem [{\citenamefont {Leadbeater}\ \emph {et~al.}(2001)\citenamefont
  {Leadbeater}, \citenamefont {Winiecki}, \citenamefont {Samuels},
  \citenamefont {Barenghi},\ and\ \citenamefont {Adams}}]{PRL.86.1410}%
  \BibitemOpen
  \bibfield  {author} {\bibinfo {author} {\bibfnamefont {M.}~\bibnamefont
  {Leadbeater}}, \bibinfo {author} {\bibfnamefont {T.}~\bibnamefont
  {Winiecki}}, \bibinfo {author} {\bibfnamefont {D.~C.}\ \bibnamefont
  {Samuels}}, \bibinfo {author} {\bibfnamefont {C.~F.}\ \bibnamefont
  {Barenghi}}, \ and\ \bibinfo {author} {\bibfnamefont {C.~S.}\ \bibnamefont
  {Adams}},\ }\href {\doibase 10.1103/PhysRevLett.86.1410} {\bibfield
  {journal} {\bibinfo  {journal} {Phys. Rev. Lett.}\ }\textbf {\bibinfo
  {volume} {86}},\ \bibinfo {pages} {1410} (\bibinfo {year}
  {2001})}\BibitemShut {NoStop}%
\bibitem [{\citenamefont {Patil}\ \emph {et~al.}(2020)\citenamefont {Patil},
  \citenamefont {Sandt}, \citenamefont {Kolle},\ and\ \citenamefont
  {Dunkel}}]{science.367.71}%
  \BibitemOpen
  \bibfield  {author} {\bibinfo {author} {\bibfnamefont {V.~P.}\ \bibnamefont
  {Patil}}, \bibinfo {author} {\bibfnamefont {J.~D.}\ \bibnamefont {Sandt}},
  \bibinfo {author} {\bibfnamefont {M.}~\bibnamefont {Kolle}}, \ and\ \bibinfo
  {author} {\bibfnamefont {J.}~\bibnamefont {Dunkel}},\ }\href {\doibase
  10.1126/science.aaz0135} {\bibfield  {journal} {\bibinfo  {journal}
  {Science}\ }\textbf {\bibinfo {volume} {367}},\ \bibinfo {pages} {71}
  (\bibinfo {year} {2020})}\BibitemShut {NoStop}%
\end{thebibliography}
%

\end{document}